\RequirePackage{fix-cm}
\documentclass[smallextended]{svjour3}       % onecolumn (second format)
\smartqed  % flush right qed marks, e.g. at end of proof
\usepackage{graphicx}
\usepackage{amsmath}
\usepackage{amssymb}
\usepackage{mathrsfs}
\usepackage{chemarrow}
\usepackage{times}

\newtheorem{thm}{Theorem}
\newtheorem{lem}[thm]{Lemma}
\newtheorem{prop}[thm]{Proposition}
\newtheorem{rem}[thm]{Remark}

\def\qed{\hfill {\vrule height5pt width5pt depth2pt}}
\begin{document}

\def\MS{{\bf S}}
\def\rd{{\rm d}}
\def\mF{{\bf F}}
\def\mQ{{\bf Q}}
\def\vJ{{\bf J}}
\def\vB{{\bf B}}
\def\bplus{{\bf +}}
\def\bplus{\mbox{\boldmath$+$}}
\def\ve{{\bf e}}
\def\pp{{\bf p}}
\def\vl{{\bf l}}
\def\vn{{\bf n}}
\def\vv{{\bf v}}
\def\vx{{\bf x}}
\def\vy{{\bf y}}
\def\vz{{\bf z}}
\def\vq{{\bf q}}
\def\vr{{\bf r}}
\def\vnu{\mbox{\boldmath$\nu$}}
\def\vxi{\mbox{\boldmath$\xi$}}
\def\vkappa{\mbox{\boldmath$\kappa$}}
\def\vgamma{\mbox{\boldmath$\gamma$}}

\title{Mathematical Formalism of Nonequilibrium Thermodynamics for Nonlinear Chemical Reaction Systems with General Rate Law%\thanks{Grants or other notes
}
%\subtitle{Nonequilibrium Thermodynamic Formalism}


\author{Hao Ge        \and
        Hong Qian %etc.
}


\institute{Hao Ge \at
              Beijing International Center for Mathematical Research (BICMR) and Biodynamic Optical Imaging Center (BIOPIC),Peking University, Beijing 100871, P.R.C. \\
              \email{haoge@pku.edu.cn}           %  \\
           \and
           Hong Qian \at
              Department of Applied Mathematics, University of Washington, Seattle, WA 98195-3925, U.S.A\\
              \email{hqian@u.washington.edu}
}

\date{Received: date / Accepted: date}

\maketitle

\begin{abstract}
This paper studies a mathematical formalism of nonequilibrium thermodynamics for chemical reaction models with $N$ species, $M$  reactions, and general rate law. We establish a mathematical basis for J. W. Gibbs' macroscopic chemical thermodynamics under G. N. Lewis' kinetic law of entire equilibrium
(detailed balance in nonlinear chemical kinetics). In doing so, the equilibrium thermodynamics is then naturally generalized to nonequilibrium settings without detailed balance.  The kinetic models are represented by a Markovian jumping process.  A generalized macroscopic chemical free energy function and its associated balance equation with nonnegative source and sink are the major discoveries. The proof is based on the large deviation principle of this type of Markov processes. A general fluctuation dissipation theorem for stochastic reaction kinetics is also proved. The mathematical theory illustrates how a novel macroscopic dynamic law can emerges from the mesoscopic kinetics in a multi-scale system.

\end{abstract}

\section{Introduction}

In recent years, the stochastic, mesosocpic nonequilibrium
thermodynamics is slowly becoming a part of the dynamic
branch of Gibbs' statistical theory of complex systems
\cite{qian-pla,qkkb}.  In addition to his statistical
ensemble theory of matters, Gibbs' another major
contribution is the macroscopic theory of chemical
thermodynamics.  He was the first to formulate a
{\em free energy function} and showed a variational
principle for chemical reactions.  Indeed,
Waage-Guldberg's law of mass action kinetics was proven
to be closely related to the variational principle with respect to
that function, connecting thermodynamics with kinetics \cite{shapiro-1965}.  This paper concerns the mathematical theory of
nonequilibrium thermodynamics for nonlinear chemical reaction
systems
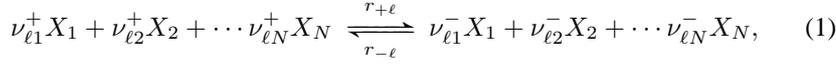
\begin{equation}
     \nu^+_{\ell 1}X_1+\nu^+_{\ell 2}X_2 + \cdots
     \nu^+_{\ell N}X_N  \  \  \underset{r_{-\ell}}{\overset{r_{+\ell}}{\rightleftharpoonsfill{26pt}}}   \  \
      \nu^-_{\ell 1}X_1+\nu^-_{\ell 2}X_2 + \cdots
     \nu^-_{\ell N}X_N,
\label{rxn}
\end{equation}
in which $1\le\ell\le M$: There are $N$ species and
$M$ reactions. $\nu_{ij}=(\nu^-_{ij}-\nu^+_{ij})$ are called
{\em stoichiometric coefficients}, they relate species to
reactions.  In a reaction vessel with rapidly stirred chemical
solutions, the concentrations of the species at time $t$,
$x_i(t)$ for $X_i$, satisfy the system of ordinary differential
equations that describes the mass balance
\begin{equation}
  \frac{\rd x_i(t)}{\rd t} = \sum_{\ell=1}^M
                \nu_{\ell i}\Big(R_{+\ell}(\vx)-R_{-\ell}(\vx)\Big),
\label{the-ode}
\end{equation}
with $\vx=(x_1,x_2,\cdots,x_N)^T$ denoting the concentrations of these chemical species, $R_{+\ell}(\vx)$ and $R_{-\ell}(\vx)$ are the forward and backward fluxes for the $\ell-$th reaction with general form. For a meaningful thermodynamic analysis, we further assume
that both $R_{\pm\ell}(\vx)\ge 0$ when all the components of $\vx$ are positive.

On the other hand, chemical reactions at the individual molecule level are
stochastic \cite{wemoerner}.
A mathematically more accurate description of the chemical
kinetics in system (\ref{rxn}) is available.
This is the stochastic theory of
Chemical Master Equation (CME) first appeared in the work of
Leontovich \cite{leontovich-35} and Delbr\"uck \cite{delbruck},
whose fluctuating trajectories can be exactly computed using the
stochastic simulation method discussed by Doob, Gillespie, and
others \cite{doob,Gillespie}. The deterministic rate equations in (\ref{the-ode}) has been proved, due to T. G. Kurtz, to be
the macroscopic large-volume limit in finite time of this stochastic chemical reaction model \cite{Kurtz78,beard-qian-book,Shwartz-Weiss-book}.  The fundamental postulate in the stochastic theory is the
notion of {\em elementary reaction}:  Each and every transition
in (\ref{rxn}) occurs with an exponential waiting time.

At finite volume $V$, denote $r_{+\ell}(\vn;V)$ and $r_{-\ell}(\vn;V)$ as the transition rates for the $\ell^{th}$ forward and backward reactions,
which satisfies $\lim_{V\rightarrow\infty}V^{-1}r_{+\ell}(V\vx;V) =R_{+\ell}(\vx)$ and $\lim_{V\rightarrow\infty}V^{-1}r_{-\ell}(V\vx;V) =R_{-\ell}(\vx)$ for any $\vx$ at the macroscopic limit $V\rightarrow\infty$ \cite{anderson-kurtz-book}.
Hence the stochastic trajectory of the copy numbers $\vn=(n_1,n_2,\cdots,n_N)$ can be expressed in terms of the
random-time-changed Poisson representation
\cite{anderson-kurtz-book}:
\begin{eqnarray}
    &&  n_j(t) \  = \  n_j(0) +
\\
  &&
\sum_{\ell=1}^M
          \nu_{\ell j}
             \left\{        Y_{+\ell}\left(
                    \int_0^t r_{+\ell}\Big(\vn(s)\Big) \rd s \right)
             -  Y_{-\ell}\left(
                    \int_0^t r_{-\ell}\Big(\vn(s)\Big) \rd s \right) \right\},
\nonumber
\end{eqnarray}
in which $Y_{+\ell}$ and $Y_{-\ell}$ are independent standard Poisson process with mean ${\bf E}(Y_{\pm\ell}(u))=u$.

The corresponding chemical master equation for the stochastic chemical reaction model, which is the Kolmogorov forward equation, is
\begin{eqnarray}
  \frac{\rd p_V(\vn,t)}{\rd t} &=& \sum_{\ell=1}^M
       \Big[   p_V(\vn-\vnu_{\ell},t)r_{+\ell}(\vn-\vnu_{\ell};V)\nonumber\\
      && - p_V(\vn,t)\Big( r_{+\ell}(\vn;V)+r_{-\ell}(\vn;V) \Big)
                 + p_V(\vn+\vnu_{\ell},t)
                    r_{-\ell}(\vn+\vnu_{\ell};V)  \Big].
\label{eq-cme}
\end{eqnarray}

	For the relatively simpler kinetics with complex balance under the law of mass action
\cite{horn-jackson,horn-1972,feinberg-1972,feinberg-91},
it has been shown recently \cite{anderson-2015}
that the steady-state large deviation rate function $\varphi^{ss}(\vx)$ for the stochastic chemical reaction model is actually a
widely used ``free energy function'' in the chemical kinetic
literature \cite{qian-beard-05} $A[\vx]=\sum_{i=1}^Nx_i\ln (x_i/x_i^{ss})-x_i+x_i^{ss}$,
where $\vx^{ss}$ is a positive stable fixed point of (\ref{the-ode}). In \cite{part1}, it was recently further shown that
$A[\vx]$ for kinetic system with complex balance,
is also an emergent quantity as the large-volume limit of the
mesoscopic free energy function $F^{(meso)}\big[p_V(\vn,t)\big]$ (See below for definition)
\cite{ge-qian-pre10}.
Hence, a macroscopic chemical thermodynamics for complex balance kinetics including detailed balance kinetics
naturally emerges.

In the present paper, we study the kinetics with general rate law with and without detailed balance.
The paper is organized as follows. Law of large number, large deviation principle and a key lemma are given in Section 2. In Section 3, we establish a mathematical basis for the macroscopic chemical thermodynamics, and naturally generalize it to general nonequilibrium settings. A general fluctuation-dissipation theorem is further proved in Section 4 for the stochastic reaction kinetics. The mathematical theory shows how, in a multi-scale system, a macroscopic dynamic law emerges from the mesoscopic kinetics at a level below.

\section{Preliminaries}

%\subsection{Mesosocopic stochastic kinetics}

\subsection{Kurtz's Law of Large Number}

Denote the right-hand-side of the rate equation (\ref{the-ode}) as $F_i(\vx)=\sum_{\ell=1}^M
                \nu_{\ell i}\Big(R_{+\ell}(\vx)-R_{-\ell}(\vx)\Big)$, and $F(\vx)=(F_1(\vx),F_2(\vx),\cdots,F_N(\vx))^T$.

The following theorem regarding the law of large number for stochastic chemical reaction models was proved by T. Kurtz in 1978 \cite{Kurtz78,Shwartz-Weiss-book}.

\begin{thm}
Assume that there exist constants $\epsilon_{\ell}$, $\Gamma$ such that
\begin{eqnarray}
&&\left|\frac{1}{V}r_{+\ell}(V\vx;V)\right|\leq \epsilon_{\ell}\big(1+|\vx|\big),~~\left|\frac{1}{V}r_{-\ell}(V\vx;V)\right|\leq \epsilon_{\ell}\big(1+|\vx|\big),
\nonumber\\
&&\left|\frac{1}{V}r_{+\ell}(V\vx;V)-R_{+\ell}(\vx)\right|\leq \frac{\Gamma\epsilon_{\ell}}{V}\big(1+|\vx|\big),
\nonumber\\
&&\left|\frac{1}{V}r_{-\ell}(V\vx;V)-R_{-\ell}(\vx)\right|\leq \frac{\Gamma\epsilon_{\ell}}{V}\big(1+|\vx|\big).
\end{eqnarray}

Denote $\vnu_\ell=(\nu_{\ell 1},\nu_{\ell 1},\cdots,\nu_{\ell N})$. Further assume
$\sum_\ell |\vnu_\ell|\epsilon_\ell<\infty$, and
there exists a positive constant $M$ such that
\begin{equation}
|F(\vx)-F(\vy)|\leq M|\vx-\vy|.
\end{equation}

Then if $\lim_{V\rightarrow\infty} \frac{\vn(0)}{V}=\vx(0)$, for any $T>0$,
\begin{equation}\label{eq-lln}
\lim_{V\rightarrow\infty}\sup_{t\leq T}\left|\frac{\vn(t)}{V}-\vx(t)\right|=0,~a.s.
\end{equation}
\end{thm}

\subsection{Large deviation principle}

Denote
$$g(\vx,{\bf \theta})= \sum_{\ell=1}^M
                 \left\{R_{+\ell}(\vx) \Big[e^{
   \vnu_{\ell}\cdot {\bf \theta}} -1\Big]+
    R_{-\ell}(\vx)\Big[e^{ -\vnu_{\ell}\cdot{\bf \theta}} -1\Big]\right\},~\theta\in \mathcal{R}^N,$$
    and
    $$l(\vx,\vy)=\sup_{\bf \theta}\left({\bf \theta}\cdot \vy-g(\vx,{\bf \theta})\right),$$
    which is called the local rate function.

Then we can define the Freidlin-Wentzell-type rate function
\begin{equation}
I_0^T(\{\vr(t):0\leq t\leq T\})=\left\{\begin{array}{ll}
\int_0^T l(\vr(s),\vr^{'}(s))ds& \text{ if } \vr(t)
 \text{ is absolutely continuous},\\[5pt]
\infty& \text{otherwise}.
\end{array}\right.
\end{equation}

The following theorem is Freidlin-Wentzell large deviation theory for stochastic chemical reaction models, which is from \cite{Shwartz-Weiss-book}.

$D^N_{[0,T]}$ is the space containing all the functions of a parameter $t\in[0,T]$ with values in $\mathcal{R}^N$ that are right continuous with left limits. Let $\Lambda$ be a collection of strictly increasing functions $\lambda$ on $[0,T]$, such that $\lambda(0)=0$ and $\lambda(T)=T$, and such that
$$\gamma(\lambda)\stackrel{\triangle}{=}\sup_{0\leq s\leq t\leq T} \left|\log\frac{\lambda(s)-\lambda(t)}{s-t}\right|<\infty.$$
The standard metric on $D^N_{[0,T]}$ is
\[
    d_d(\{\vx(t):0\leq t\leq T\},\{\vy(t):0\leq t\leq T\})
       \hspace{2in}
\]
\[
  =\inf_{\lambda\in\Lambda}\left\{\max\left(\gamma(\lambda),\sup_{0\leq t\leq T}|x(t)-y(\lambda(t))|\right)\right\}.
\]
$(D^N_{[0,T]},d_d)$ is a complete, severable metric space. It is called the Skorohod space \cite{Shwartz-Weiss-book}.

\begin{thm}\label{thm-ldp}
Assume for each $\ell$, $\log R_{+\ell}(\vx)$ and $\log R_{-\ell}(\vx)$ are bounded and Lipschitz continuous. Then $I_0^T$ is a good rate function in $(D^N_{[0,T]},d_d)$, and\\
(i) For every closed set $F\in D^N_{[0,T]}$ and every $\vx$,
\begin{eqnarray}
&&\limsup_{V\rightarrow\infty} \frac{1}{V}\log P_{\vx}\left(\left\{\frac{\vn(t)}{V}:0\leq t\leq T\right\}\in F\right)\nonumber\\
&\leq& -\inf\left\{I_0^T(\{\vr(t):0\leq t\leq T\}):\{\vr(t):0\leq t\leq T\}\in F,\vr(0)=\vx\right\},
\end{eqnarray}
(ii) For every open set $G\in D^N_{[0,T]}$, uniformly for $\vx$ in compact sets,
\begin{eqnarray}
&&\liminf_{V\rightarrow\infty} \frac{1}{V}\log P_{\vx}\left(\left\{\frac{\vn(t)}{V}:0\leq t\leq T\right\}\in G\right)\nonumber\\
&\geq& -\inf\left\{I_0^T(\{\vr(t):0\leq t\leq T\}):\{\vr(t):0\leq t\leq T\}\in G,\vr(0)=\vx\right\}.
\end{eqnarray}
\end{thm}

%\begin{rem}
%Typically, what we usually use is the case of 'open sausage'
%$$G_\epsilon=\left\{\{z(t):0\leq t\leq T\}\in D^N_{[0,T]}:\sup_{0\leq t\leq T}|r(t)-z(t)|<\epsilon\right\},$$
%and closed sausage
%$$F_\epsilon=\left\{\{z(t):0\leq t\leq T\}\in D^N_{[0,T]}:\sup_{0\leq t\leq T}|r(t)-z(t)|\leq\epsilon\right\}.$$
%\end{rem}

The following theorem is also from \cite{Shwartz-Weiss-book}, following the Freidlin-Wentzell large deviation theory for stochastic chemical reaction models.

\begin{thm}
Assume\\
 (a) $\vq$ is the global attractive unique stable fixed point of the rate equation (\ref{the-ode});\\
 (b) for each $\ell$, $\log R_{+\ell}(\vx)$ and $\log R_{-\ell}(\vx)$ are bounded and Lipschitz continuous in some neighborhood of $\vq$;\\
 (c) for each $V$, the stochastic chemical reaction models are positive recurrent with steady state probability $p^{ss}_V$ for the stochastic process $\vn(t)$.

Then for any $\epsilon$,
 \begin{equation}\label{eq-focus}
\lim_{V\rightarrow\infty}\pi_V(B_\epsilon({\vq}))=1,
\end{equation}
in which $B_\epsilon({\vq})$ is the $\epsilon$-neighborhood of the position $\vq$ in $R^N$, and $\pi_V(B_\epsilon)$ is defined as $\pi_V(B_\epsilon)=\sum_{\frac{\vn}{V}\in B_\epsilon}p^{ss}_V(\vn)$.

Let $D\subset \mathcal{R}^N$ be a smooth, bounded open set. Define
$$S\stackrel{\triangle}{=}\left\{\{\vr(t)\}:r(0)=\vq,r(T)\in \bar{D}~for~some~T>0\right\}.$$
Define
$$I^*\stackrel{\triangle}{=}\inf\left\{I_0^T(\{\vr(t):0\leq t\leq T\}):\{\vr(t):0\leq t\leq T\}\in S\right\}.$$

Further assume\\
 (d) for each $\delta>0$, there is a $T<\infty$ such that, uniformly over $\vx_0\in B_\epsilon(\vq)$, with $\vx(0)=\vx_0$, such that
 $$|\vx(t)-\vq|<\delta, \forall t>T.$$
 (e) $S$ is a continuity set, and every point in $S$ is the limit of points in the interior of $S$;\\
 (f) there is some $\epsilon_0$ so that $D\subset B_{\epsilon_0}(\vx)$ and $\varphi^{ss}(\vy)>I^*(S)+1$ whenever $\vy$ is outside $B_{\epsilon_0}(\vx)$.

Then
\begin{equation}
\lim_{V\rightarrow\infty}\frac{1}{V} \log\pi_V(D)=-\inf_{\vx\in D}\varphi^{ss}(\vx),
\label{eq-ldp}
\end{equation}
in which
$$\varphi^{ss}(\vx)=\inf_{t\geq 0}\inf_{r(0)=\vq,r(t)=\vx}\{I_0^t(\{r(s):0\leq s\leq t\})\}.$$
\end{thm}

\begin{rem}
Applying the contraction principle, one can straightforwardly prove that the distribution of $\frac{\vn(t)}{V}$ as a function of $V$ satisfies the large deviation principle with a good rate function $\varphi_t(\vx)$, for any given $t$. Also, one can use the techniques in Chapter 6 of \cite{Freidlin} to generalize (\ref{eq-ldp}) to the case with multiple attractors.
\end{rem}

Next we would like to derive the differential equation that $\varphi^{ss}(\vx)$ satisfies. We first need a lemma.

\begin{lem}\label{key-lemma}
For each $\epsilon$, assume $k^{\epsilon}(V,\vx)=\pi_{V}(B_{\epsilon}(\vx))\exp\{V\inf_{\vy\in B_{\epsilon}(\vx)}\varphi^{ss}(\vy)\}$ is continuous, and there exists another positive continuous function $f^{\epsilon}(V)$ of $V$ such that
$$\lim_{V\rightarrow\infty,\epsilon\rightarrow 0^+}\frac{k^{\epsilon}(V,\vx)}{f^{\epsilon}(V)}=K(\vx)$$
exists, and uniformly at any sufficiently small neighborhood of each $\vx$. We also assume that both the function $K(\vx)$ and $\varphi^{ss}(\vx)$ are at least twice continuous differentiable and $K(\vx)$ is positive.

Then for any positive function $\tilde{\epsilon}(V)$ such that $V\tilde{\epsilon}(V)\geq \frac{1}{2}$ and
$$\lim_{V\rightarrow\infty}V\tilde{\epsilon}^2(V)=0,$$
we have $\pi_V(B_{\tilde{\epsilon}(V)}(\vx))>0$ and
\begin{equation}
\lim_{V\rightarrow +\infty}\frac{\pi_V(B_{\tilde{\epsilon}(V)}(\vx-\vnu/V))}{\pi_V(B_{\tilde{\epsilon}(V)}(\vx))}=e^{\vnu\cdot\nabla_{\vx}\varphi^{ss}(\vx)},\label{lem5_eq1}
\end{equation}
for any $N$-dimensional integer vector $\vnu$. The convergence is uniform in any sufficient small neighborhood of $\vx$.

Furthermore, denote $\vn(\vx,V)$ as the nearest integer vector to the point $\vx V$, then
\begin{equation}
\lim_{V\rightarrow\infty}\frac{p^{ss}_V(\vn(\vx,V)-\vnu)}{p^{ss}_V(\vn(\vx,V))}=e^{\vnu\cdot\nabla_{\vx}\varphi^{ss}(\vx)}.\nonumber
\end{equation}
\end{lem}

{\bf Proof:}

%According to Theorem \ref{thm-ldp}, we know that $\lim_{V\rightarrow\infty}\frac{1}{V}\log k^{\epsilon}(V,\vx)=0$.

%Also assume the limit is uniform in any bounded open set inside $D_{\vx}^{\epsilon}$.

According the assumptions, we know that for each $\vx$, given any $\delta>0$, there exists a constant $V_0$ and $\epsilon_0$, such that once $V>V_0$ and $\epsilon<\epsilon_0$,
$$\left|\frac{k^{\epsilon}(V,\vx)}{f^{\epsilon}(V)}-K(\vx)\right|<\delta/3,$$
$$\left|\frac{k^{\epsilon}(V,\vx-\vnu/V)}{f^{\epsilon}(V)}-K(\vx-\vnu/V)\right|<\delta/3.$$

Combined with the fact that for sufficiently large $V$
$$\left|K(\vx)-K(\vx-\vnu/V)\right|<\delta/3,$$
therefore we have
$$\lim_{V\rightarrow\infty,\epsilon\rightarrow 0^+}\frac{k^{\epsilon}(V,\vx-\vnu/V)}{f^{\epsilon}(V)}=K(\vx),$$
followed by
$$\lim_{V\rightarrow\infty,\epsilon\rightarrow 0^+}\frac{k^{\epsilon}(V,\vx-\vnu/V)}{k^{\epsilon}(V,\vx)}=1.$$

Noticing that
\begin{eqnarray}
\pi_V(B_{\tilde{\epsilon}(V)}(\vx-\vnu/V))&=&k^{\epsilon}(V,\vx-\vnu/V)\exp\{-V\inf_{\vy\in B_{\tilde{\epsilon}(V)}(\vx-\vnu/V)}\varphi^{ss}(\vy)\},\nonumber\\
\pi_V(B_{\tilde{\epsilon}(V)}(\vx))&=&k^{\epsilon}(V,\vx)\exp\{-V\inf_{\vy\in B_{\tilde{\epsilon}(V)}(\vx)}\varphi^{ss}(\vy)\},\nonumber
\end{eqnarray}
we only need to prove
$$\lim_{V\rightarrow\infty}V\left[\inf_{\vy\in B_{\tilde{\epsilon}(V)}(\vx)}\varphi^{ss}(\vy)-\inf_{\vy\in B_{\tilde{\epsilon}(V)}(\vx-\vnu/V)}\varphi^{ss}(\vy)\right]=\vnu\cdot\nabla_{\vx}\varphi^{ss}(\vx).$$

%estimate the difference between $V\left\{\inf_{\vy\in D_{\vx}^{\epsilon}}\varphi^{ss}(\vy)\}-\inf_{\vy\in D_{\vx-\vnu/V}^{\epsilon}}\varphi^{ss}(\vy)\}\right\}$ and $-V\left[\varphi^{ss}(\vx-\vnu/V)-\varphi^{ss}(\vx)\right]$.

Since the function $\varphi^{ss}(\vx)$ is at least twice continuously differentiable, then when $V$ is large, the first and second derivatives of  $\varphi^{ss}(\vx)$ inside $B_{\tilde{\epsilon}(V)}(\vx)\cup B_{\tilde{\epsilon}(V)}(\vx-\vnu/V)$ are bounded. Hence
$$\inf_{\vy\in B_{\tilde{\epsilon}(V)}(\vx-\vnu/V)}\varphi^{ss}(\vy)-\varphi^{ss}(\vx-\vnu/V)=-\sum_{i=1}^N \left|\frac{\partial \varphi^{ss}(\vx-\vnu/V)}{\partial x_i}\right|\tilde{\epsilon}(V)+O(\tilde{\epsilon}^2(V)),$$
and
$$\inf_{\vy\in B_{\tilde{\epsilon}(V)}(\vx)}\varphi^{ss}(\vy)-\varphi^{ss}(\vx)=-\sum_{i=1}^N \left|\frac{\partial \varphi^{ss}(\vx)}{\partial x_i}\right|\tilde{\epsilon}(V)+O(\tilde{\epsilon}^2(V)).$$

Therefore
\begin{eqnarray}
&&\left|\inf_{\vy\in B_{\tilde{\epsilon}(V)}(\vx-\vnu/V)}\varphi^{ss}(\vy)-\inf_{\vy\in B_{\tilde{\epsilon}(V)}(\vx)}\varphi^{ss}(\vy)+\varphi^{ss}(\vx)-\varphi^{ss}(\vx-\vnu/V)\right|\nonumber\\
&=&\left|\sum_{i=1}^N \left[\left|\frac{\partial \varphi^{ss}(\vx)}{\partial x_i}\right|-\left|\frac{\partial \varphi^{ss}(\vx-\vnu/V)}{\partial x_i}\right|\right]\tilde{\epsilon}(V)\right|+O(\tilde{\epsilon}^2(V))\nonumber\\
&=&O(\frac{1}{V}\tilde{\epsilon}(V))+O(\tilde{\epsilon}^2(V))=o(\frac{1}{V}),
\end{eqnarray}
and the fact
$$\lim_{V\rightarrow\infty} V\left[\varphi^{ss}(\vx)-\varphi^{ss}(\vx-\vnu/V)\right]=\vnu\cdot\nabla_{\vx}\varphi^{ss}(\vx)$$
concludes the proof of Eq. \ref{lem5_eq1}.

Once $\tilde{\epsilon}(V))=\frac{1}{2V}$, $\pi^{ss}_V(B_{\tilde{\epsilon}(V)}(\vx))=p^{ss}_V(\vn(\vx,V))$ and $\pi_V(B_{\tilde{\epsilon}(V)}(\vx-\vnu/V))=p_V(\vn(\vx,V)-\vnu)$.

\qed

Denote the stoichiometric matrix $\mathcal{S}_{N\times M}$, in which $s_{ij}=\nu_{ji}$. Any vector in the left null space sets a conservation law to the chemical reaction system, i.e. once any vector ${\bf \eta}=(\eta_1,\eta_2,\cdots,\eta_N)$ satisfies
$$\sum_{i=1}^N \eta_is_{ij}=0,~\forall j,$$
then
$$\frac{d\left(\sum_{i=1}^N\eta_ix_i\right)}{dt}=0.$$

Therefore, the surviving space of the reaction scheme (\ref{rxn}), no matter stochastic or deterministic, can be described by $\mathcal{L}^{\vx}=\{\vx+\sum_{\ell=1}^M \xi_{\ell}\vnu_{\ell}^T, \forall \vxi=(\xi_1,\xi_2,\cdots,\xi_M)\}$ (called the surviving space indicated by $\vx$), in which $\vx$ is any given state in $\mathcal{R}^N$.

\begin{thm}\label{thm-quasi-potential}
Under the assumptions in Theorem \ref{thm-ldp} and Lemma \ref{key-lemma}, and further assume that the limit
$$\lim_{V\rightarrow\infty}\frac{r_{\pm\ell}(V\vx;V)}{V}=R_{\pm\ell}(\vx)$$
is locally uniform for any sufficiently small neighborhood of each $\vx$.

Then the function $\varphi^{ss}(\vx)$, also called {\bf quasi potential}, satisfies
\begin{equation}
  \sum_{\ell=1}^M
                 R_{+\ell}(\vx) \Big[1 - e^{
   \vnu_{\ell}\cdot\nabla_{\vx}\varphi^{ss}(\vx)} \Big]+
    R_{-\ell}(\vx)\Big[1-  e^{ -\vnu_{\ell}\cdot\nabla_{\vx}\varphi^{ss}(\vx)} \Big]
     = 0,
\label{hge}
\end{equation}
i.e.
$$g(\vx,\nabla_{\vx}\varphi^{ss}(\vx))=0.$$
Then $\varphi^{ss}(\vx)$ satisfies $\frac{\rd}{\rd t}\varphi^{ss}\big(\vx(t)\big) \le 0$, and the equality holds if and only if the gradient of $\varphi^{ss}(\vx)$ in the surviving space $\mathcal{L}^{\vx}$ indicated by $\vx$ vanishes at $\vx$, i.e. $\vnu_{\ell}\cdot\nabla_{\vx}\varphi^{ss}(\vx)=0$ for all $\ell$.

Furthermore, the steady state condition at $\vx$, i.e.
$$F_i(\vx)=\sum_{\ell=1}^M
                \nu_{\ell i}\Big(R_{+\ell}(\vx)-R_{-\ell}(\vx)\Big)=0,~\forall i$$
is a sufficient condition for the vanishing of the gradient of $\varphi^{ss}(\vx)$ in $\mathcal{L}^{\vx}$; while if $\varphi^{ss}(\vx)$ is twice differentiable and the dimension of the Hessian matrix of $\varphi^{ss}(\vx)$ is equal to the dimension of the surviving space, i.e the column space of the stoichiometric matrix $\mathcal{S}$, then the steady state condition is also a necessary condition for the vanishing of the gradient of $\varphi^{ss}(\vx)$ in $\mathcal{L}^{\vx}$.
\end{thm}

{\bf Proof:}
Given $V$, denote $\vn(\vx,V)$ as the nearest integer vector to the point $\vx V$. Then the chemical master equation (\ref{eq-cme}) at steady state can be rewritten as
\begin{eqnarray}
\sum_{\ell=1}^M &&\frac{p^{ss}_V(\vn(\vx,V)-\vnu_{\ell})}{p^{ss}_V(\vn(\vx,V))}r_{+\ell}(\vn(\vx,V)-\vnu_{\ell};V)- r_{+\ell}(\vn(\vx,V);V)\nonumber\\
       && + \frac{p^{ss}_V(\vn(\vx,V)+\vnu_{\ell})}{p^{ss}_V(\vn(\vx,V))}r_{-\ell}(\vn(\vx,V)+\vnu_{\ell};V)- r_{-\ell}(\vn(\vx,V);V)=0.
\end{eqnarray}

Then according to Lemma \ref{key-lemma}, we have
\begin{eqnarray}
\lim_{V\rightarrow\infty}\frac{p^{ss}_V(\vn(\vx,V)-\vnu_{\ell})}{p^{ss}_V(\vn(\vx,V))}=e^{\vnu_{\ell}\cdot\nabla_{\vx}\varphi^{ss}(\vx)};\nonumber\\
\lim_{V\rightarrow\infty}\frac{p^{ss}_V(\vn(\vx,V)+\vnu_{\ell})}{p^{ss}_V(\vn(\vx,V))}=e^{-\vnu_{\ell}\cdot\nabla_{\vx}\varphi^{ss}(\vx)}.
\end{eqnarray}

Also similar to the proof of Lemma \ref{key-lemma}, we know that
\begin{eqnarray}
\lim_{V\rightarrow\infty}\frac{r_{+\ell}(\vn(\vx,V)-\vnu_{\ell};V)}{V}=\lim_{V\rightarrow\infty}\frac{r_{+\ell}(\vn(\vx,V);V)}{V}=R_{+\ell}(\vx);\nonumber\\
\lim_{V\rightarrow\infty}\frac{r_{-\ell}(\vn(\vx,V)+\vnu_{\ell};V)}{V}=\lim_{V\rightarrow\infty}\frac{r_{-\ell}(\vn(\vx,V);V)}{V}=R_{-\ell}(\vx),\nonumber\\
\end{eqnarray}
which proved the Eq. \ref{hge}.

Finally since $e^{\vx}\geq \vx+1$
\begin{eqnarray}
  \frac{\rd}{\rd t}\varphi^{ss}\big(\vx(t)\big) &=&
              \frac{\rd\vx(t)}{\rd t} \cdot
              \nabla_{\vx}\varphi^{ss}(\vx)
\nonumber\\
	&=& \sum_{\ell=1}^M \Big(R_{+\ell}(\vx)
              -R_{-\ell}(\vx)\Big)\vnu_{\ell}\cdot
               \nabla_{\vx}\varphi^{ss}(\vx)
\nonumber\\
	&\le& \sum_{\ell=1}^M
                 R_{+\ell}(\vx) \Big[e^{
   \vnu_{\ell}\cdot\nabla_{\vx}\varphi^{ss}(\vx)}-1\Big]+
    R_{-\ell}(\vx)\Big[e^{ -\vnu_{\ell}\cdot\nabla_{\vx}\varphi^{ss}(\vx)}-1\Big]\nonumber\\
     &=& 0.\label{neq-lyapunov}
\end{eqnarray}

The equality holds if and only if $\vnu_{\ell}\cdot\nabla_{\vx}\varphi^{ss}(\vx)=0$ for all $\ell$.

The surviving space $\mathcal{L}^{\vx}$ indicated by $\vx$ is $\{\vx+\sum_{\ell=1}^M \xi_{\ell}\vnu_{\ell}^T, \forall \xi\}$. Therefore, the gradient of $\varphi^{ss}(\vx)$ in $\mathcal{L}^{\vx}$ with respect to $\xi$
$$\frac{\partial \varphi^{ss}(\vx+\sum_{\ell=1}^M \xi_{\ell}\vnu_{\ell}^T)}{\partial \xi_{\ell}}|_{\vxi=0}=\vnu_{\ell}\cdot\nabla_{\vx}\varphi^{ss}(\vx).$$

At steady state $F(\vx)=0$, resulting in $\frac{\rd}{\rd t}\varphi^{ss}\big(\vx(t)\big)=F^T(\vx)\cdot \nabla_{\vx}\varphi^{ss}(\vx)=0$, i.e. the equality in (\ref{neq-lyapunov}) holds.

%disproof

%If the second derivative of $\varphi^{ss}(\vx)$ along each direction does not vanish and $\vx$ is not the steady state, then $F(\vx^)$ doesnot vanish. We already know that
%$$F(\vx^{'})\cdot \nabla_{\vx}\varphi^{ss}\big(\vx^{'}\big)\leq 0,$$
%for each $\vx^{'}$, hence we know that in a sufficient small neighborhood of $\vx$, the second derivative of $\varphi^{ss}(\vx)$ along the direction of $F(\vx)$ must be zero, which %contradicts with the assumption.

On the other hand, take the derivative of the left-hand-side of Eq. \ref{hge}, we can get
\begin{eqnarray}
\sum_{\ell=1}^M&& \left\{ \frac{\partial R_{+\ell}(\vx)}{\partial \vx_k} \Big[1 - e^{
   \vnu_{\ell}\cdot\nabla_{\vx}\varphi^{ss}(\vx)} \Big]+
    \frac{\partial R_{-\ell}(\vx)}{\partial \vx_k}\Big[1-  e^{ -\vnu_{\ell}\cdot\nabla_{\vx}\varphi^{ss}(\vx)} \Big]\right\}\nonumber\\
   &&+\sum_{\ell=1}^M\left[R_{-\ell}(\vx)e^{ -\vnu_{\ell}\cdot\nabla_{\vx}\varphi^{ss}(\vx)}-R_{+\ell}(\vx)e^{
   \vnu_{\ell}\cdot\nabla_{\vx}\varphi^{ss}(\vx)}\right]\sum_{1\leq i\leq N} \nu_{\ell i}\frac{\partial^2 \varphi^{ss}(\vx)}{\partial \vx_i\partial \vx_k}\nonumber\\
     =&& 0,
\end{eqnarray}
for each $1\leq k\leq N$.

Once $\vnu_{\ell}\cdot\nabla_{\vx}\varphi^{ss}(\vx)=0$ for each $\ell$, the above equation becomes
\begin{equation}
\sum_{1\leq i\leq N}F_i(\vx)\frac{\partial^2 \varphi^{ss}(\vx)}{\partial \vx_i\partial \vx_k}=0,
\end{equation}
for each $1\leq k\leq N$.

Hence if the dimension of the Hessian matrix $\left\{\frac{\partial^2 \varphi^{ss}(\vx)}{\partial \vx_i\partial \vx_j}\right\}$ is the same as the dimension of the stoichiometric matrix $\mathcal{S}$, and notice that $F(\vx)$ is also in the column space of $\mathcal{S}$, we know that $F(\vx)=0$.

\qed

\begin{rem}
(1) The equation (\ref{hge}) has already known to the community of physics and chemistry for a long time, dated back to Gang Hu's work in early 1980s\cite{hugang,galstyan-saakian,anderson-2015}. It is also known as Eikonal equation as well as Hamilton-Jacobi equation (HJE) for the stochastic chemical reaction models.
Indeed, Theorem \ref{thm-quasi-potential} can also be proved through the approach of classic mechanics \cite{Arnold-book,Landau-book}, regarding the $I_0^t(\{r(s):0\leq s\leq t\})$ as the action functional of the trajectory $\{r(s):0\leq s\leq t\}$. Such a proof exists and has already been exposed in detail by Feng and Kurtz \cite{Feng_book}. Their approach started from a Hamiltonian-Lagrangian convergence formalism, which considers the convergence of the semigroup
$$L(t) f(\vx)\stackrel{\triangle}{=}\frac{1}{V}\log E\left[e^{Vf(\frac{\vn(t)}{V})}|\frac{\vn(0)}{V}=\vx\right].$$

(2) The solution of the partial differential equation (\ref{hge}) is not unique. To make sense of a physical solution of it, one should turn to the weak KAM theory, especially the representation formula of Eikonal equations \cite{Fathi2005,Ishii2007}.

(3) It is known that the solutions of (\ref{hge}) are generally only Lipschitz and semi-convex. There are points of Lebesgue measure zero that are non-differentiable even for first order derivatives. In the present paper, we avoid getting involved in such a subtle and complicated situation, hence we simply assume at least twice differentiability of the solution.
\end{rem}

\subsection{Diffusion approximation}\label{sec-diffusion}

Diffusion approximation to the chemical reaction models can be achieved either through the Kramers-Moyal expansion of the chemical master equation or Poisson representation \cite{Kampen,Keizer,Kurtz78}. The approximated  diffusion process $\vz(t)$ is a Ito process with drift coefficient $F(\vz)$ and the matrix of diffusion coefficient $A(\vz)$, in which
$$A_{ij}(\vz)=\frac{1}{V}\sum_{\ell=1}^M (R_{+\ell}(\vz)+R_{-\ell}(\vz))\nu_{\ell i}\nu_{\ell_j},~1\leq i,j\leq N.$$
Kurtz has proved that the difference at finite time interval between $\vx(t)$ and $\vz(t)$ is at the order of $\frac{\log V}{\sqrt{V}}$ \cite{Kurtz78}.

The diffusion process close to any stable fixed point $\vq$ of Eq. \ref{the-ode} can be further approximated by a linear diffusion process with the constant diffusion matrix $A=A(\vq)$ and a linear drift coefficient $B\vz$, in which
$$B_{ij}=\frac{\partial F_i}{\partial \vx_j}(\vq ).$$

Denote $\Xi$ as the variance of the invariant measure of such a linear diffusion process, then we have
\begin{equation}
A=-(B\Xi+\Xi B).
\end{equation}
It is referred to as fluctuation-dissipation theorem of the chemical reaction model by Keizer \cite{Keizer}. We will give a clear mathematical statement of this theorem, especially what $\Xi$ is for the original stochastic chemical reaction model, as well as a rigorous proof in Section \ref{sec-FDT}.

\section{Nonequilibrium thermodynamic formalism}

\subsection{Mesoscopic stochastic thermodynamics}

Recently, a framework of nonequilibrium thermodynamics for mesoscopic stochastic dynamics has been proposed\cite{ge-qian-pre10,esposito2007,esposito-vandenbroeck}. In the stochastic chemical reaction model, the entropy production at time $t$ is defined as
\begin{eqnarray}
    e_p\big[p_V(\vn,t)\big] &=& \sum_{\ell=1}^M \sum_{\vn}
                  \Big(p_V(\vn,t)r_{+\ell}(\vn;V)-
             p_V(\vn+\vnu_{\ell},t) r_{-\ell}(\vn+\vnu_{\ell};V) \Big)
\nonumber\\
	&& \times\ln\left(\frac{p_V(\vn,t)r_{+\ell}(\vn;V)}{p_V(\vn+\vnu_{\ell},t) r_{-\ell}(\vn+\vnu_{\ell};V) }\right),
\end{eqnarray}
which can be decomposed into two nonnegative terms, i.e.
$$e_p\big[p_V(\vn,t)\big] = f_d\big[p_V(\vn,t)\big]  + Q_{hk}\big[p_V(\vn,t)\big],$$
in which
\begin{eqnarray}
	f_d\big[p_V(\vn,t)\big] &=& \sum_{\ell=1}^M \sum_{\vn}
                 \Big(p_V(\vn,t)r_{+\ell}(\vn;V)-
             p_V(\vn+\vnu_{\ell},t) r_{-\ell}(\vn+\vnu_{\ell};V) \Big)
\nonumber\\
	&& \times\ln\left(\frac{p_V(\vn,t)p^{ss}(\vn+\vnu_{\ell}) }{p_V^{ss}(\vn)p_V(\vn+\vnu_{\ell},t) }\right)
\end{eqnarray}
is the free energy dissipation rate, while
\begin{eqnarray}
	Q_{hk}\big[p_V(\vn,t)\big] &=& \sum_{\ell=1}^M \sum_{\vn}
                       \Big(p_V(\vn,t)r_{+\ell}(\vn;V)-
             p_V(\vn+\vnu_{\ell},t) r_{-\ell}(\vn+\vnu_{\ell};V) \Big)
\nonumber\\
	&& \times\ln\left(\frac{p_V^{ss}(\vn)r_{+\ell}(\vn;V)}{p_V^{ss}(\vn+\vnu_{\ell}) r_{-\ell}(\vn+\vnu_{\ell};V) }\right)
\end{eqnarray}
is the dissipation rate of housekeeping heat that represents the external driving force \cite{HS2001,ge-qian-pre13}.

All three $e_p\big[p_V(\vn,t)\big] $, $f_d\big[p_V(\vn,t)\big] $ and $Q_{hk}\big[p_V(\vn,t)\big]$ are nonnegative \cite{ge-qian-pre10,esposito2007,esposito-vandenbroeck}.  Furthermore, the $f_d[p_V(\vn,t)]$ is actually the time derivative of a mesoscopic general free energy $F^{meso}[p_V(\vn,t)]$, i.e.
$$f_d[p_V(\vn,t)]=-\frac{\rd}{\rd t}F^{meso}[p_V(\vn,t)],$$
where
\begin{equation}
    F^{(meso)}\big[p_V(\vn,t)\big] =
       \sum_{\vn} p_V(\vn,t)\ln\left(\frac{p_V(\vn,t)}{
                      p^{ss}(\vn)}\right).
\label{meso-F}
\end{equation}

$f_d\big[p_V(\vn,t)\big]=0$ if and only if the process is at steady state, i.e. $p_V(\vn,t)\equiv p_V^{ss}(\vn)$. And $Q_{hk}\big[p_V(\vn,t)\big]=0$ if and only if the detailed balance condition is satisfied at steady state, i.e. $p_V^{ss}(\vn)r_{+\ell}(\vn;V)=p_V^{ss}(\vn+\vnu_{\ell}) r_{-\ell}(\vn+\vnu_{\ell};V)$. Noticing that the steady-state distribution $p_V^{ss}(\vn)$ only depends on the transition rates $r_{+\ell}(\vn;V)$ and $r_{-\ell}(\vn;V)$, not related to the transient distribution at time $t$, i.e. $p_V(\vn,t)$, we know that $Q_{hk}\big[p_V(\vn,t)\big]>0$ implies that there is certain external force driving the process, which inputs free energy into the system and dissipates heat.

\subsection{Macroscopic limits}

The following theorem proves a widely hold believe in chemical and
biochemical kinetics
\cite{beard-qian-PLoS1}.

\begin{thm}
Assume the Eqn. \ref{eq-lln},\ref{eq-focus}\ref{eq-ldp},\ref{hge}, as well as the assumptions of Lemma \ref{key-lemma} hold, and also assume that the limit
$$\lim_{V\rightarrow\infty}\frac{r_{\pm\ell}(V\vx;V)}{V}=R_{\pm\ell}(\vx)$$
is locally uniform for any sufficiently small neighborhood of each $\vx$.

Then denote $\vx(t)$ as the solution to Eq. \ref{the-ode},
as $V$ goes to infinity, we have for each time $t$

(a)
\begin{equation}
\lim_{V\rightarrow\infty}\frac{e_p\big[p_V(\vn,t)\big]}{V}=\sigma^{(tot)}\big[\vx(t)\big],
\end{equation}
where the density of macroscopic chemical
entropy production rate
\begin{equation}
	\sigma^{(tot)}[\vx] = \sum_{\ell=1}^M \Big(R_{+\ell}(\vx)
       -R_{-\ell}(\vx)\Big)
            \ln\left(\frac{R_{+\ell}(\vx)}{
               R_{-\ell}(\vx) }\right);\nonumber
\end{equation}

(b)
\begin{equation}
\lim_{V\rightarrow\infty}\frac{f_d\big[p_V(\vn,t)\big]}{V}=f_d^{(macro)}[\vx(t)],
\end{equation}
where the density of macroscopic free energy dissipation rate
\begin{equation}
    f_d^{(macro)}[\vx] = \sum_{\ell=1}^M
                  \Big(R_{-\ell}(\vx)-R_{+\ell}(\vx) \Big)
          \vnu_{\ell}\cdot
       \nabla_{\vx} \varphi^{ss}(\vx).\nonumber
\end{equation}

(c)
\begin{equation}
\lim_{V\rightarrow\infty}\frac{F^{(meso)}\big[p_V(\vn,t)\big]}{V}=\varphi^{ss}(\vx(t)),~\frac{d\varphi^{ss}(\vx)}{dt}=-f_d^{(macro)}[\vx],
\end{equation}
hence $\varphi^{ss}(\vx)$ can be regarded as the density of general macroscopic free energy;

(d)
\begin{equation}
\lim_{V\rightarrow\infty}\frac{Q_{hk}\big[p_V(\vn,t)\big]}{V}=q_{hk}^{(macro)}[\vx(t)],
\end{equation}
where the density of macroscopic housekeeping heat dissipation rate
\begin{equation}
    q_{hk}^{(macro)}[\vx] =  \sum_{\ell=1}^M \Big(R_{-\ell}(\vx)-R_{+\ell}(\vx) \Big)
            \ln\left(\frac{R_{-\ell}(\vx)}{
               R_{+\ell}(\vx) }\ e^{ -\vnu_{\ell}\cdot\nabla_{\vx}\varphi^{ss}(\vx)} \right).\nonumber
\end{equation}
\end{thm}

\begin{rem}
We here only need the validity of the law of large number and large deviation theory, not the technical assumptions under which they can be proved so far in the previous section. Hence the results can possibly hold for much more general settings.
\end{rem}

{\bf Proof: }

According to the strong law of large number (Eq. \ref{eq-lln}), we know that given any small $\epsilon>0$, for sufficiently large $V$, the probability concentrates on the integers satisfying $\frac{\vn}{V}\in B_\epsilon(\vx(t))$.

Given any small positive number $\delta$, for those $\vn$, when $V$ is sufficiently large,

$$\left|\frac{r_{+\ell}(\vn;V)}{V}-R_{+\ell}(\frac{\vn}{V})\right|<\delta,$$

and since $\left|R_{+\ell}(\frac{\vn}{V})-R_{+\ell}(\vx)\right|<C_{+\ell}(\vx)\epsilon$ for some constant $C_{+\ell}(\vx)$, we have

$$\left|\frac{r_{+\ell}(\vn;V)}{V}-R_{+\ell}(\vx)\right|<\delta+C_{+\ell}(\vx)\epsilon.$$

Similarly

$$\left|\frac{r_{-\ell}(\vn;V)}{V}-R_{-\ell}(\vx)\right|<\delta+C_{-\ell}(\vx),$$

and

$$\left|\ln\left(\frac{r_{+\ell}(\vn;V)}{r_{-\ell}(\vn+\vnu_{\ell};V) }\right)-\ln\frac{R_{+\ell}(\vx)}{R_{-\ell}(\vx)}\right|<\delta+C_1(\vx)\epsilon.$$

Then, based on the large deviation principle (Eq. \ref{eq-ldp}) and the assumption of Lemma \ref{key-lemma}, we know that for sufficiently large $V$, once $\frac{\vn}{V}\in B_\epsilon(\vx(t))$,

$$\left|\frac{1}{V}\ln p^{ss}_V(\vn)+\inf_{\vy\in B_{\frac{1}{2V}}(\frac{\vn}{V})}\varphi^{ss}(\vy)\right|<\delta,$$

followed by

$$\left|\frac{1}{V}\ln p^{ss}_V(\vn)+\varphi^{ss}(\vx)\right|<\delta+C_2(\vx)\epsilon,$$

since
$$\left|\inf_{\vy\in B_{\frac{1}{2V}}(\frac{\vn}{V})}\varphi^{ss}(\vy)-\varphi^{ss}(\vx)\right|<C_2(\vx)\epsilon.$$

Furthermore, due to Lemma \ref{key-lemma}, for sufficiently large $V$, once $\frac{\vn}{V}\in B_\epsilon(\vx(t))$,

$$\left|\ln\frac{p^{ss}_V(\vn)}{p^{ss}_V(\vn+\vnu_{\ell})}-\vnu_{\ell}\cdot \nabla_{\vx}\varphi^{ss}(\frac{\vn}{V})\right|<\delta,$$

and since

$$\left|\vnu_{\ell}\cdot \nabla_{\vx}\varphi^{ss}(\frac{\vn}{V})-\vnu_{\ell}\cdot \nabla_{\vx}\varphi^{ss}(\vx)\right|<C_3(\vx)\epsilon,$$

it follows with
$$\left|\ln\frac{p^{ss}_V(\vn)}{p^{ss}_V(\vn+\vnu_{\ell})}-\vnu_{\ell}\cdot \nabla_{\vx}\varphi^{ss}(\vx)\right|<C_3(\vx)\epsilon.$$

Finally, similar to the proof of Lemma \ref{key-lemma} based on the large deviation principle for $\{p_V(\vn,t)\}$ given $t$, we have

$$\left|\frac{1}{V}\ln p_V(\vn,t)+\varphi_t(\vx(t))\right|<\delta+C_4(\vx)\epsilon,$$

and

$$\left|\ln\frac{p_V(\vn,t)}{p_V(\vn+\vnu_{\ell},t)}-\vnu_{\ell}\cdot \nabla_{\vx}\varphi_t(\vx(t))\right|<\delta+C_5(\vx)\epsilon.$$

Noticing the large deviation rate function $\varphi_t(\vx)$ for the transient distributions at time $t$ taking its global minimum at $\vx=\vx(t)$, i.e. $\varphi_t(\vx(t))=\nabla_{\vx}\varphi_t(\vx(t))=0$, we know

$$\left|\frac{1}{V}\ln p_V(\vn,t)\right|<\delta+C_4(\vx)\epsilon,$$

and

$$\left|\ln\frac{p_V(\vn,t)}{p_V(\vn+\vnu_{\ell},t)}\right|<\delta+C_5(\vx)\epsilon.$$

Putting all these together, and realizing that $\epsilon$ and $\delta$ can be arbitrarily small, we finally prove (a)-(d).

\qed

\begin{prop}\label{Prop_three_terms}
The three terms $\sigma^{(tot)}\big[\vx\big]$, $f_d^{(macro)}[\vx]$ and $q_{hk}^{(macro)}[\vx]$ are all nonnegative. And

(a) $\sigma^{(tot)}\big[\vx\big]=0$ if and only if the {\bf strong} detailed balance condition is satisfied at $\vx$:
$$R_{+\ell}(\vx)=R_{-\ell}(\vx),~\forall \ell.$$

(b) $q_{hk}^{(macro)}[\vx]=0$ if and only if the {\bf weak} detailed balance condition is satisfied at $\vx$:
\begin{equation}
     \ln\left(\frac{R_{+\ell}(\vx)}{R_{-\ell}(\vx)}\right)
  = -\vnu_{\ell}\cdot\varphi^{ss}(\vx)),~\forall \ell.
\label{wdbc}
\end{equation}

(c) $f_d^{(macro)}[\vx]=0$ if and only if the following condition is satisfied, i.e.
$$\vnu_{\ell}\cdot \nabla_{\vx} \varphi^{ss}(\vx)=0,~\forall \ell,$$
which is equivalent to the vanishing of the gradient of $\varphi^{ss}(\vx)$ in the surviving space $\mathcal{L}^{\vx}$ indicated by $\vx$. If the dimension of the Hessian matrix of $\varphi^{ss}(\vx)$ is equal to the dimension of the surviving space, i.e the column space of the stoichiometric matrix $\mathcal{S}$, then it is also equivalent to the steady state condition
$$F_i(\vx)=\sum_{\ell=1}^M
                \nu_{\ell i}\Big(R_{+\ell}(\vx)-R_{-\ell}(\vx)\Big)=0,~\forall i.$$
\end{prop}

{\bf Proof:}

(a) is straightforward.

(b) Note that $\forall x\geq 0$, $\ln x\geq 1-\frac{1}{x}$
and $\ln x\leq x-1$.  Thus we have
\begin{subequations}
\begin{equation}
  \ln\left(\frac{R_{-\ell}(\vx)}{
               R_{+\ell}(\vx) }e^{-\vnu_{\ell}\cdot
       \nabla_{\vx} \varphi^{ss}(\vx)}\right)
         \geq 1-\left(\frac{R_{+\ell}(\vx) }{R_{-\ell}(\vx)}\right)
                e^{\vnu_{\ell}\cdot
       \nabla_{\vx} \varphi^{ss}(\vx)},
\end{equation}
and
\begin{equation}
  \ln\left(\frac{R_{-\ell}(\vx)}{
               R_{+\ell}(\vx) }e^{-\vnu_{\ell}\cdot
       \nabla_{\vx} \varphi^{ss}(\vx)}\right)
      \leq \left(\frac{R_{-\ell}(\vx)}{
               R_{+\ell}(\vx) }\right)
            e^{-\vnu_{\ell}\cdot
       \nabla_{\vx} \varphi^{ss}(\vx)}-1.
\end{equation}
\label{2-ineq}
\end{subequations}
Therefore
\begin{eqnarray}
  q^{(macro)}_{hk}\big[\vx\big]
    &\geq& \sum_{\ell=1}^M \left\{ R_{-\ell}(\vx)\left[1-\left(
             \frac{R_{+\ell}(\vx) }{R_{-\ell}(\vx)}\right)
          e^{\vnu_{\ell}\cdot
       \nabla_{\vx} \varphi^{ss}(\vx)}\right] \right.
\nonumber\\
&& \left. +R_{+\ell}(\vx)
      \left[1-\left(\frac{R_{-\ell}(\vx)}{
               R_{+\ell}(\vx) }\right) e^{-\vnu_{\ell}\cdot
       \nabla_{\vx} \varphi^{ss}(\vx)}\right]   \right\}
         \ =  \  0,
\end{eqnarray}
and the equality holds if and only if Eq. (\ref{wdbc}) is satisfied.

(c) Note that $\forall x\in \mathcal{R}$, $\vx\geq 1-e^{-\vx}$. Thus we have
\begin{eqnarray}
    f_d^{(macro)} &= &\sum_{\ell=1}^M
                  \Big(R_{-\ell}(\vx)-R_{+\ell}(\vx) \Big)
          \vnu_{\ell}\cdot \nabla_{\vx} \varphi^{ss}(\vx)\nonumber\\
          &\geq& \sum_{\ell=1}^M R_{-\ell}(\vx)\left[1-\exp\{-\vnu_{\ell}\cdot \nabla_{\vx} \varphi^{ss}(\vx)\}\right]\nonumber\\
          &&+R_{+\ell}(\vx)\left[1-\exp\{\vnu_{\ell}\cdot \nabla_{\vx} \varphi^{ss}(\vx)\}\right]=0,
\end{eqnarray}
and the equality holds if and only if $\vnu_{\ell}\cdot \nabla_{\vx} \varphi^{ss}(\vx)=0$ for all $\ell$. Then applying the Theorem \ref{thm-quasi-potential}.

\qed

%\subsection{A balance equation for the macroscopic chemical energy $\varphi^{ss}(\vx)$}

	We thus have derived a macroscopic law of chemical (free)
energy balance:
\begin{equation}
   \frac{\rd}{\rd t}\varphi^{ss}\big(\vx(t)\big) =
           q^{(macro)}_{hk}\big[\vx(t)\big]
            - \sigma^{(tot)}\big[\vx(t)\big],
\label{febe}
\end{equation}
in which all three macroscopic {\em densities}, including the
free energy dissipation rate,
$f_d^{(macro)}\big[\vx(t)\big]\equiv-\frac{\rd}{\rd t}\varphi^{ss}\big(\vx(t)\big)$,
house-keeping heat rate,
$q_{hk}^{(macro)}\big[\vx(t)\big]$ and
total entropy production rate
$\sigma^{(tot)}\big[\vx(t)\big]$, are non-negative.
$q_{hk}^{(macro)}$ and $\sigma^{(tot)}$ should be
identified as the ``source'' and the ``sink'' for an emergent,
macroscopic chemical energy function $\varphi^{ss}$.
It is clear
that $\varphi^{ss}$ is a consequence of a global,
infinitely long time behavior of the mesoscopic system.

\section{Fluctuation-dissipation theorem} \label{sec-FDT}

Here we further give a rigorous statement and proof for the fluctuation-dissipation theorem for general stochastic chemical reaction models, which is first discovered by
Joel Keizer in 1980s through diffusion approximation \cite{Keizer}.

\begin{thm}
$\vq$ is any stable fixed point of Eq. \ref{the-ode}. Assume $\varphi^{ss}(\vx)$ is at least twice differentiable. Define $\Xi_{ij}=\frac{\partial^2 \varphi^{ss}(\vq)}{\partial \vx_i\partial \vx_j}$, $A_{ij}=\sum_{\ell=1}^M(R_{+\ell}(\vq)+R_{-\ell}(\vq))\nu_{\ell i}\nu_{\ell j}$, and $B_{ij}=\frac{\partial F_i(\vq)}{\partial \vx_j}$. Then we have the equality
\begin{equation}
\Xi A\Xi=-\Xi B-B\Xi.
\end{equation}
If $\Xi$ is invertible, then $A=-\left(B\Xi^{-1}+\Xi^{-1}B\right)$.
\end{thm}

{\bf Proof: }
Take the second derivative of the left-hand-side of Eq. \ref{hge} and notice $\vnu_{\ell}\cdot\nabla_{\vx}\varphi^{ss}(\vq)=F_i(\vq)=0$ for each $\ell$ and $i$, then we can get
\begin{eqnarray}
 &&\sum_{\ell=1}^M \left\{\left[-\frac{\partial R_{+\ell}(\vq)}{\partial \vx_k}+\frac{\partial R_{-\ell}(\vq)}{\partial \vx_k}\right ]\sum_{1\leq i\leq N} \nu_{\ell i}\frac{\partial^2 \varphi^{ss}(\vq)}{\partial \vx_i\partial \vx_j} \right\}\nonumber\\
   &+&\sum_{\ell=1}^M \left\{\left[-\frac{\partial R_{+\ell}(\vq)}{\partial \vx_j}+\frac{\partial R_{-\ell}(\vq)}{\partial \vx_j}\right ]\sum_{1\leq i\leq N} \nu_{\ell i}\frac{\partial^2 \varphi^{ss}(\vq)}{\partial \vx_i\partial \vx_k} \right\}\nonumber\\
   &-&\sum_{\ell=1}^M(R_{+\ell}(\vq)+R_{-\ell}(\vq))\sum_{1\leq i\leq N} \nu_{\ell i}\frac{\partial^2 \varphi^{ss}(\vq)}{\partial \vx_i\partial \vx_j}\sum_{1\leq i\leq N} \nu_{\ell i}\frac{\partial^2 \varphi^{ss}(\vq)}{\partial \vx_i\partial \vx_k}=0,
\end{eqnarray}
for each $1\leq j,k\leq N$. It is exactly the equality $\Xi A\Xi=-\Xi B-B\Xi$.

\qed

\begin{rem}
1. For the diffusion approximation close to any stable fixed point $\vq$ in Section \ref{sec-diffusion}, the diffusion matrix is $\frac{A(\vq)}{V}$ and the variance of the stationary linear diffusion is just $\frac{\Xi^{-1}}{V}$. Now $\Xi$ has a more rigorous and clear definition from the original stochastic chemical reaction.

2. The diffusion approximation violates the conservation laws, while the chemical reaction models keep it.
\end{rem}

\section{Kinetics with detailed balance}

	Wegscheider-Lewis cycle condition and detailed
balance are cornerstone concepts of chemical
kinetics and equilibrium thermodynamics
\cite{gnlewis,shapiro-1965,shear-jtb,shear-jcp,schuster-schuster}.

Let $\vxi=(\xi_1,\cdots,\xi_M)$ be a $M$-dimensional
vector and
\[
           \sum_{\ell=1}^M \xi_\ell\nu_{\ell i} = 0,~\forall i=1,2,\cdots,N.
\]
That is $\vxi$ is in the right null space of stoichiometric
matrix $\mathcal{S}_{N\times M}$, in which $s_{ij}=\nu_{ji}$.  If for every vector $\vxi$
in the right null space of $\mathcal{S}$ one has
\begin{equation}
   \sum_{\ell=1}^M \xi_{\ell}
      \ln\left(\frac{R_{+\ell}(\vx)}{R_{-\ell}(\vx)}\right)
    = 0,  \   \forall \vx;
\label{0033}
\end{equation}
then we say the kinetics satisfies Wegscheider-Lewis cycle condition
\cite{gnlewis,shapiro-1965,shear-jtb,shear-jcp,schuster-schuster},
or loop law \cite{qian-beard-liang,qian-beard-05}.

\begin{prop}
	For the chemical reaction models, the following statements are equivalent:

($i$)  The chemical reaction model satisfies the Wegscheider-Lewis cycle condition;

($ii$)  The weak detailed balance condition (\ref{wdbc}) is satisfied for $\forall\vx,\ell$;

($iii$)  The macroscopic house-keeping heat dissipation rate $q^{(macro)}_{hk}[\vx]=0$ $\forall\vx$;

($iv$) The time-evolution of the general free energy function
\begin{equation}
      \frac{\rd\varphi^{ss}[\vx(t)]}{\rd t}
      = -\sigma^{(tot)}[\vx(t)].
\end{equation}

Furthermore, (iv) also implies any stable steady state of the macroscopic kinetics (\ref{the-ode})
$\vx^{ss}$ satisfies the strong detailed balance condition.
\end{prop}

{\bf Proof:}

($i$) $\rightarrow$ ($ii$): Due to the Wegscheider-Lewis cycle condition, we can choose the rates $r_{+\ell}(\vn;V)$ for the forward reaction and $r_{-\ell}(\vn;V)$ for the backward reaction, such that for sufficiently large $V$, the stochastic chemical reaction model satisfies the Kolmogorov cycle condition, i.e. for each possible cycle in the state space, the multiply of reaction rates for the forward cycle and the multiply of reaction rates for the backward cycle is equal to each other  \cite{JQQ}. The strategy can be first set $r_{\pm\ell}(\vn;V)=VR_{\pm\ell}(\frac{\vn}{V})$, and then fine tuning it sequentially following the elementary cycles with smallest numbers of states.

Kolmogorov cycle condition is equivalent to the stationary distribution satisfying
detailed balance
\begin{equation}
     \frac{p_V^{ss}(\vn+\vnu_{\ell}) }{p_V^{ss}(\vn) }
      = \frac{r_{+\ell}(\vn;V)}{r_{-\ell}(\vn+\vnu_{\ell};V)}.
\end{equation}
Therefore in the macroscopic limit, it yields based on Lemma \ref{key-lemma}
\begin{equation}
  -\vnu_{\ell}\cdot
            \nabla_{\vx} \varphi^{ss}(\vx)
      = \ln\left(\frac{R_{+\ell}(\vx)}{R_{-\ell}(\vx)}\right).
\label{0024}
\end{equation}

Noticing that the definition of $\varphi^{ss}(\vx)$ is only dependent on $R_{\pm\ell}(\vx)$, as the limit of $r_{\pm\ell}(V\vx;V)/V$, the weak detailed balance condition holds independent of the choosing of $r_{\pm\ell}(\vn;V)$.

($ii$) $\rightarrow$ ($i$): straightforward.

($ii$)  $\Leftrightarrow$ ($iii$):
Already proved in the Proposition \ref{Prop_three_terms}.

($iii$) $\Leftrightarrow$ ($iv$): Obviously if we notice that $\sigma^{(tot)}[\vx]=f_d^{macro}[\vx]$ if and only if $q_{hk}^{macro}[\vx]=0$.

\qed

For the special case of Waage and Guldberg's law of mass
action:
\begin{equation}
           R_{+\ell}(\vx)= k_{+\ell}\prod_{j=1}^N x_j^{\nu^+_{\ell j}},
      \  \
          R_{-\ell}(\vx)= k_{-\ell}\prod_{j=1}^N x_j^{\nu^-_{\ell j}}.
\label{R4lma}
\end{equation}

For reaction kinetics with
mass-action rate law and weak detailed balance for each $\vx$, $\vnu_{\ell}\cdot\nabla_{\vx}\varphi^{ss}(\vx)=$
$\ln(k_{-\ell}/k_{+\ell})+\vnu_{\ell}\ln\vx$ $=\Delta\mu_{\ell}$, the chemical potential {\em difference} of the $\ell^{th}$ reaction.
Then one also has $\vnu_{\ell}\cdot\nabla_{\vx}\varphi^{ss}(\vx)=$ $\vnu_{\ell}\ln(\vx/\vx^{ss})$ \cite{anderson-2015}. Therefore,
$\varphi^{ss}(\vx)$
is precisely the macroscopic Gibbs chemical potential
$\mu(\vx)$. It is now worth going back
to take a fresh look how Gibbs was able to connect his two
major accomplishments, statistical mechanics and
macroscopic chemical thermodynamics, using the idea
of ``variational method for virtual change'' to show the
Lyapunov function.  This is {\em a fundamental
theorem of nonequilibrium chemical thermodynamics}.

\begin{rem}
For general rate law, weak detailed balance satisfied by each $\vx$ might not
always guarantees a unique steady state $x^{ss}$ to (\ref{the-ode}).
The following condition is necessary if a steady state is
unique \cite{othmer}:
\begin{equation}
    \vnu_{\ell}\cdot\nabla_{\vx}\ln\left(\frac{R_{+\ell}(\vx)}{
                   R_{-\ell}(\vx)}\right) < 0
\end{equation}
for all $\vx$ and $\ell$.

From Eq. \ref{0024}, we know that once the weak detailed balance condition is satisfied,
\begin{equation}
  -\sum_{k,j=1}^N \nu_{\ell j}
            \left(\frac{\partial^2\varphi^{ss}(\vx)}{
         \partial x_j\partial x_k}\right)\nu_{\ell k}
      =  \vnu_{\ell}\cdot \nabla_{\vx}
         \ln\left(\frac{R_{+\ell}(\vx)}{R_{-\ell}(\vx)}\right),
\label{eq028}
\end{equation}
hence if
$\varphi^{ss}(\vx)$ is convex, then the left-hand-side of (\ref{eq028}) is negative
for all $\vnu_{\ell}$ and $\vx>0$. So the $\vx^{ss}$ is
unique.

	For Waage and Guldberg's law of mass action with (\ref{R4lma}), we have
\begin{equation}
 \vnu_{\ell}\cdot\nabla_{\vx}\ln\left(\frac{R_{+\ell}(\vx)}{
                   R_{-\ell}(\vx)}\right)  = \sum_{j=1}^N
             \nu_{\ell j} \frac{\partial}{\partial x_j}
             \sum_{k=1}^N \ln x_k^{-\nu_{\ell k}}
       = -\sum_{j=1}^N
             \frac{\nu_{\ell j}^2}{x_j}<0,
\end{equation}
hence in this case, weak detailed balance can guarantee the uniqueness of $\vx^{ss}$.
\end{rem}

\begin{rem}
In the case of law of mass action, strong detailed balance condition at steady state $\vx^{ss}$ also implies the weak detailed balance for each $\vx$. It is because in this case, the quasi potential $\varphi^{ss}(\vx)$ can be explicitly written down (see the following section) \cite{anderson-2015}.
\end{rem}

\section{Kinetics with complex balance}

	The stationary state in reaction systems with strong detailed
balance is called an equilibrium steady state.  Not every
stationary state in mesoscopic stochastic reaction
systems (\ref{rxn}) with kinetics (\ref{eq-cme}) is
an equilibrium state; a nonequilibrium steady state
(NESS) is characterized by a non-zero entropy
production rate $\sigma^{tot}\big[\vx^{ss}\big]=$
$q_{hk}^{macro}\big[\vx^{ss}\big]>0$:
The amount of energy required to sustain the NESS
is precisely balanced by the amount of entropy
production.

Noting Eq. (\ref{hge}) can be re-written as
\begin{eqnarray}
   && \sum_{\ell=1}^M
                \Big[ R_{+\ell}(\vx)
          -R_{-\ell}(\vx) e^{ -\vnu_{\ell}\cdot\nabla_{\vx}\varphi^{ss}(\vx)} \Big]
\Big[1
       -e^{
   \vnu_{\ell}\cdot\nabla_{\vx}\varphi^{ss}(\vx)} \Big] = 0,
\label{0039}
\end{eqnarray}
hence the validity of this equation at NESS is not due to zero for each and every $\ell$ term, but a
balance between terms of different $\ell$s.

	If $R_{\pm\ell}(\vx)$ are given by the law of mass action
(\ref{R4lma}), and introducing
\begin{equation}
    A\big[\vx\big] = \sum_{j=1}^N \left[ x_j
             \ln \left(\frac{x_j}{x_j^{ss}}\right) - x_j + x_j^{ss}\right].
\label{Afunction}
\end{equation}
Then
$\nabla_{\vx} A[\vx]=\ln (\vx/\vx^{ss})$, and
\begin{equation}
  \vnu_{\ell}\cdot\nabla_{\vx} A[\vx]=
          \sum_{j=1}^N \nu_{\ell j}\ln \left(\frac{x_j}{x_j^{ss}}\right)
            = \ln \prod_{j=1}^N
                 \left(\frac{x_j}{x_j^{ss}}\right)^{\nu_{\ell j}}
              = \ln\left(\frac{R_{-\ell}(\vx)R_{+\ell}(\vx^{ss})}{
           R_{+\ell}(\vx)R_{-\ell}(\vx^{ss})}\right).
\label{ggg}
\end{equation}
Therefore, the left-hand-side of (\ref{0039}) is
\begin{eqnarray}
  && \sum_{\ell=1}^M  \Big(R_{+\ell}(\vx^{ss})-R_{-\ell}(\vx^{ss})\Big)
       \left(\frac{R_{+\ell}(\vx)}{R_{+\ell}(\vx^{ss})}
   -\frac{R_{-\ell}(\vx)}{R_{-\ell}(\vx^{ss})}
             \right)
\nonumber\\
	&=&  \sum_{\ell=1}^M  \Big(R_{+\ell}(\vx^{ss})-R_{-\ell}(\vx^{ss})\Big)
      \left[ \prod_{j=1}^N
         \left(\frac{x_j}{x_j^{ss}} \right)^{\nu^+_{\ell j}}-
           \prod_{j=1}^N
         \left(\frac{x_j}{x_j^{ss}} \right)^{\nu^-_{\ell j}} \right].
\label{0041}
\end{eqnarray}

A mass-action kinetics is called complex balanced if one of its steady
state $\vx^{ss}$ guarantees the right-hand-side of (\ref{0041}) being zero
for all $\vx$ \cite{horn-jackson,horn-1972,feinberg-1972,feinberg-91}.  In other words, under the law of mass action
and assuming $\vx^{ss}$ is one of its steady state, the
$A[\vx]$ defined in (\ref{Afunction}) is a solution to Eq. \ref{hge} for kinetics with complex balance.  We further note
that $A[\vx]$ is convex, and the macroscopic kinetics (\ref{the-ode}) always goes downhill along $A[\vx]$.  Therefore, the
$\vx^{ss}$ is the unique steady state.

In fact, Anderson et al. have
shown that $\varphi^{ss}(\vx)=A[\vx]$ in the system with law of mass action and complex balance.
\cite{anderson-2015}. Thus,
\begin{subequations}
\begin{eqnarray}
    f_d^{(macro)}[\vx] &=&-\frac{\rd  \varphi^{ss}\big[\vx\big] }{
          \rd t}  \ =  -\sum_{i=1}^N \frac{\rd x_i}{\rd t}
               \ln\left(\frac{x_i}{x_i^{ss}}\right)
\nonumber\\
	&=& \sum_{\ell=1}^M \sum_{i=1}^N
                 \nu_{\ell i}\Big(R_{-\ell}[\vx]-R_{+\ell}[\vx]\Big)
                      \ln\left(\frac{x_i}{x_i^{ss}}\right)
\nonumber\\
	&=& \sum_{\ell=1}^M \Big(R_{-\ell}[\vx]-R_{+\ell}[\vx]\Big)
            \ln\left(\frac{R_{-\ell}[\vx]R_{+\ell}[\vx^{ss}]}{
                     R_{+\ell}[\vx]R_{-\ell}[\vx^{ss}]}\right),
\\
	q_{hk}^{(macro)}[\vx] &=& \sigma^{(tot)}[\vx]-
                 f^{(macro)}_d[\vx]
\nonumber\\
	&=&  \sum_{\ell=1}^M \Big(R_{+\ell}[\vx]-R_{-\ell}[\vx]\Big)
            \ln\left(\frac{R_{+\ell}[\vx^{ss}]}{
                     R_{-\ell}[\vx^{ss}]}\right).
\end{eqnarray}
\end{subequations}
These two quantities were introduced in \cite{part1}
based purely on chemical thermodynamic intuition.
It is now shown to be ``derivable'' from the general
law of macroscopic chemical energy balance, Eq. \ref{febe}.

\begin{acknowledgements}
The authors would like to thank Xiao Jin, Tiejun Li for comments and helpful discussions. H. Ge
is supported by NSFC (No. 21373021 and 11622101), and the 863 program (No. 2015AA020406).
\end{acknowledgements}



\end{document}